\documentclass[sigconf]{acmart}  
\copyrightyear{2022}
\acmYear{2022}
\setcopyright{rightsretained}
\acmConference[FAccT '22]{2022 ACM Conference on Fairness, Accountability, and Transparency}{June 21--24, 2022}{Seoul, Republic of Korea}
\acmBooktitle{2022 ACM Conference on Fairness, Accountability, and Transparency (FAccT '22), June 21--24, 2022, Seoul, Republic of Korea}\acmDOI{10.1145/3531146.3534647}
\acmISBN{978-1-4503-9352-2/22/06}

\usepackage{graphicx}
\usepackage[shortlabels]{enumitem}
\usepackage{tabularx}
\usepackage{xcolor}
\usepackage{hyperref}

\usepackage[multiple]{footmisc}

\usepackage[caption=false]{subfig}

\begin{document}
\title[CrowdWorkSheets]{CrowdWorkSheets: Accounting for Individual and Collective Identities Underlying Crowdsourced Dataset Annotation}

\author{Mark Díaz}
\email{markdiaz@google.com}
\orcid{0000-0003-0167-9839}
\affiliation{
  \institution{Google Research}
  \city{New York}
  \state{New York}
  \country{USA}
}

\author{Ian D.\ Kivlichan}
\email{kivlichan@google.com}
\orcid{0000-0003-2719-2500}
\affiliation{
  \institution{Jigsaw, Google}
  \city{New York}
  \state{New York}
  \country{USA}
}
\author{Rachel Rosen}
\email{rachelrosen@google.com}
\orcid{0000-0003-2927-1245}
\affiliation{
  \institution{Jigsaw, Google}
  \city{New York}
  \state{New York}
  \country{USA}
}

\author{Dylan K. Baker}
\email{dylan@dair-institute.org}
\orcid{0000-0002-2468-2347}
\affiliation{
  \institution{Distributed AI Research Institute}
  \city{Seattle}
  \state{Washington}
  \country{USA}
}
\authornote{Work completed while at Google.}

\author{Razvan Amironesei}
\email{amironesei@gmail.com}
\orcid{0000-0002-3497-0641}
\affiliation{
  \institution{Google Research}
  \city{Des Moines}
  \state{Iowa}
  \country{USA}
}

\author{Vinodkumar Prabhakaran}
\email{vinodkpg@google.com}
\orcid{0000-0003-3329-2305}
\affiliation{
  \institution{Google Research}
  \city{San Francisco}
  \state{California}
  \country{USA}
}

\author{Emily Denton}
\email{dentone@google.com}
\orcid{0000-0003-4915-0512}
\affiliation{
  \institution{Google Research}
  \city{New York}
  \state{New York}
  \country{USA}
}

%
\renewcommand{\shortauthors}{D\'iaz et al.}

\begin{abstract}
Human annotated data plays a crucial role in machine learning (ML) research and development. However, the ethical considerations around the processes and decisions that go into dataset annotation have not received nearly enough attention. In this paper, we survey an array of literature that provides insights into ethical considerations around crowdsourced dataset annotation. We synthesize these insights, and lay out the challenges in this space along two layers: (1) who the annotator is, and how the annotators' lived experiences can impact their annotations, and (2) the relationship between the annotators and the crowdsourcing platforms, and what that relationship affords them. Finally, we introduce a novel framework, CrowdWorkSheets, for dataset developers to facilitate transparent documentation of key decisions points at various stages of the data annotation pipeline: task formulation, selection of annotators, platform and infrastructure choices, dataset analysis and evaluation, and dataset release and maintenance.
\end{abstract}


%
\settopmatter{printfolios=true}

\maketitle

\section{Introduction}
Human computation refers to the practice of tapping into human intelligence and cognition as computational elements within an information processing system design, often done on a large global scale \cite{quinn2011human}. The sheer scale of human computation that the Web enables has made possible things that were previously unimaginable, e.g., \textit{Captchas} digitizing the entire NYTimes historical publications, global participatory platforms for human rights and crises response,\footnote{\url{https://www.ushahidi.com/}} and large-scale data and distributed analyses enabled by citizen science projects.\footnote{\url{https://www.citizenscience.gov/}} In particular, human computation has played a critical role in the research, development, and deployment of modern-day artificial intelligence systems, through the creation of training datasets \cite{imagenet_cvpr09}, and human-in-the-loop systems \cite{Filippova2019, monarch2021human}. By enabling efficient and scalable distribution of data labelling microtasks, crowdsourcing platforms are a natural choice for dataset developers aiming to cheaply and efficiently generate dataset annotations. 

In this paper we explore the challenges and decision points inherent to crowdsourced annotation of machine learning datasets and propose a framework, CrowdWorkSheets, for reflecting on data-set annotation decisions, and documenting them in a standardized manner. At a high level, CrowdWorkSheets prompts dataset developers to ask: who is annotating the data, and why is that important? We consider how the ethical concerns of data annotation intersect with the identities of the annotators, the social structures surrounding their work, and how their individual perspectives may become encoded within the dataset labels. In doing so, we push back against the prevalent notion that crowdworkers are interchangeable and instead seek to illuminate why they are not. Data generated in crowdwork tasks is shaped by a range of social factors and the datasets that workers help to build continue to shape systems long after worker engagement ends. Processes of annotation thus impact future models built from this data; therefore, understanding the perspectives captured through data labeling is crucial to fully understanding these models and the potential social impact they can have.

Our work is motivated by, and extends, prior scholarship examining ethical considerations relating to crowdsourcing. For instance, \citet{vakharia2015beyond} outline various kinds of challenges encountered in this space by analyzing and comparing seven different crowdsourcing platforms. In addition, \citet{schlagwein2019ethical} conducted extensive fieldwork, engaging crowdworkers, platform organizers, and requesters over the course of three years to uncover a range of ethical dilemmas relating to gig economy crowdsourcing. \citet{shmueli2021beyond} identified risks of harm to crowdworkers engaged in NLP tasks. Attending to ethical issues more broadly, \citet{kocsis2016towards} used value-sensitive design and transparency literature to develop a taxonomic framework of ethical considerations in crowdsourcing.

Our primary contribution is the introduction of CrowdWorkSheets, a novel framework designed to facilitate critical reflection and transparent documentation of dataset annotation decisions, processes, and outcomes. CrowdWorkSheets complements and extends dataset development and documentation frameworks that have previously been developed in service of transparency, accountability, and reproducibility \citep{Gebru2018, bender2018data, Holland2018, chmielinski2020DNP, hutchinson2021accountability,ramirez2021state,datacards}, but focuses specifically on unique considerations relating to crowdsourced data-set annotation. Similar to recent dataset documentation frameworks that have tailored to specific domains (e.g. \citep{artsheets, healthsheets}), our work starts from a recognition of the limitations of ``one-size-fits-all'' solutions to ethical issues in dataset development. More specifically, we offer CrowdWorkSheets as a targeted intervention to address unresolved ethical problems in crowdsourcing that relate specifically to worker subjectivity and worker experiences.
 
 The remainder of this paper is structured as follows. First, re review literature relating to (1) how annotators' individual and collective social experiences can impact their annotations, and (2) the relationship between the annotators and the crowdsourcing platforms, and what that relationship means for their ability to engage in fair work. Next, we introduce the CrowdWorkSheets considerations and documentation questions. 
 Finally, we step through a hypothetical case study to illustrate how a dataset developer might use CrowdWorkSheets to document their decisions.

\section{Who is annotating ML datasets and why does it matter?}
\label{sec_challenges}
The historical lineage of crowdsourced labor can be traced back to manufacturing innovation of piecework \citep{Alkhatib2017}---a form of labor that produced the ``unskilled worker'' as the paradigmatic \textit{interchangeable} component and which has been credited with giving rise to the productivity and ingenuity of American manufacturing \citep{freeman2018}. In an analogous manner, crowdwork platforms are often designed to position crowdworkers as \textit{interchangeable} \citep{irani2013turkopticon}.
While some forms of digital work can be decomposed and distributed, the presumption that crowdsourced dataset annotators exercise near-identical capacities of perception and judgement ignores 
the fact that social position, identity, and experience shape how annotators apply knowledge.
Yet, recent empirical work has revealed that dataset annotators are often treated as interchangeable in practice. For example, relatively little attention is given or documented about annotator positionality---how annotator social identity shapes their understanding of the world \citep{geiger2020, scheuerman2021}. Crowd workers are often selected by task requesters based on quality metrics, rather than on any socially defining features of their knowledge or experience. This is concerning; when crowd-sourced annotations are used to build datasets capturing subjective phenomena, such as sentiment or hate speech, annotators' values and subjective judgments shape the perspectives that machine learning models learn from in a manner that is wholly unaccounted for. 

\subsection{Accounting for the socio-cultural backgrounds of annotators}
Understanding  socio-cultural factors of an annotator pool---or even selecting annotators based on these factors---is important because annotator's identity and lived experience can impact how annotation questions are interpreted and responded to. More generally, subjective interpretations of a task can produce divergent annotations across different communities \citep{sen2015turkers}. As \citet{aroyo2015truth} argue, the notion of ``one truth'' in crowdsourcing responses is a myth; disagreement between annotators, which is often viewed as problematic noise, can actually provide a valuable signal.

A variety of social, cultural, economic, and infrastructural factors contribute to the sociodemographic distribution of workers on any given platform. 
For example, as \citep{gray2019ghost} points out, the remote nature of crowdwork differentially attracts workers along gender lines, such as mothers who do crowdwork because it allows for an easier balance of childcare in comparison to other work. 
Other work similarly notes significant gender differences among workers who report engaging in crowdwork because they are only able to conduct work from their homes \citep{berg2015income}. This leads to a different gender balance among crowdworkers in the United States than in many other parts of the world; crowdworkers in most of the world are disproportionately male, while in contrast over 60\% of U.S. annotators are female  \cite{posch2018characterizing}. \citet{ipeirotis2010demographics} hypothesizes this to be due to the remote nature of the work, which attracts stay-at-home parents and unemployed or underemployed adults, who are more likely to be women. Additionally, health problems and disability are also a factor that cause many workers to only be able to work from home and motivates them to pursue crowd work \citep{berg2015income}.

Since many crowdsourced annotator pools are sociodemographically skewed, there are implications for which populations and cultural values are represented in datasets and models \cite{ghosh-etal-2021-detecting} as well as which populations face the challenges of crowdwork \citep{irani2013turkopticon, gray2019ghost}. 
Accounting for skews in annotator demographics is critical for contextualizing datasets and ensuring responsible downstream use. In short, there is value in acknowledging, and accounting for, worker's socio-cultural background---both from the perspective of data quality and societal impact.

\subsection{Lived experiences of annotators as expertise}
Just as substantive work experience lends valuable domain expertise for a given problem (e.g., annotation of medical imagery by a medical professional), lived experience with, and proximity to, a problem domain can provide a valuable source of expertise for dataset annotation. For example, women experience higher rates of sexual harassment online compared to men, and among those who have experienced online abuse, women are more likely to identify it as such \citep{vogels2021state}. This underscores the importance of considering raters' experience with gender-based harassment, when using crowdwork to annotate/moderate online harassment. Recent work has highlighted how the ``average'' rater, in terms of gender and other social characteristics, varies dramatically depending on which geographies raters are selected from \cite{posch2018characterizing}. Additionally, as a result of the previously mentioned sociodemographic differences among who is likely to conduct crowdwork, ratings on sexual harassment data, for example, may differ according to the geographic distribution of raters. 

At the same time, relevant lived experience among annotators does not always fall along demographic lines. \citet{waseem2016racist} demonstrated that incorporating feminist and antiracist activists' perspectives into hate speech annotations yielded better aligned models. Similarly, \citet{patton2019annotating} demonstrated the importance of situated domain expertise---including contextualized knowledge of local language, concepts, and gang activity---when annotating Twitter images to detect pathways to violence among gang-involved youth in Chicago. They found that expert annotations (i.e., those from individuals situated in Chicago and with community ties) significantly diverged from those of graduate students who were scholars of social work and who were trained to perform the annotation task but who lacked this lived experience.

In summary, a core question to answer in data collection is how much annotator's identity, lived experience, and prior knowledge of a problem space matters for the task at hand, and how it impacts what the resulting dataset is intended to capture. While the aforementioned examples constitute relatively subjective tasks, even seemingly objective tasks such as annotating medical text vary surprisingly with annotator backgrounds and experience. \citet{aroyo2015truth} show that medical experts are more likely to erroneously identify medical relations as being expressed in text compared with non experts because the experts already know the relation is true based on knowledge external to the task. Their work underscores a need to examine annotator experience even in tasks that appear to be unambiguous or objective.

\section{Worker Experiences of Dataset Annotation}\label{experiences}

Another series of considerations are rooted in annotators' experiences with annotation work itself and how those experiences impact how they do their work. These include issues related to worker compensation, power imbalances in between worker and requester, and the structure of annotation work itself---all of which can pose barriers to crowdworker well-being and their ability to produce quality work.

\subsection{Compensation and working conditions}
Compensation policies of crowdwork platforms should be a core aspect to consider when thinking about responsible data collection. For instance, in the U.S., there are currently no regulations around worker pay for crowdwork \citep{berg2015income}, and the Fair Labor Standards Act that established the minimum wage,\footnote{\url{https://www.dol.gov/agencies/whd/flsa}} is not applicable for crowdworkers as they are independent contractors \citep{semuels2018internet}. Reports on how much crowdworkers actually earn vary, but generally show an average lower than minimum wage \cite {irani2013turkopticon}; surveys of workers from Amazon Mechanical Turk and Crowdflower place it on average between \$1 and \$5.5 per hour \cite{berg2015income} with a median wage of roughly \$2 / hour  \cite{hara2018data,semuels2018internet}; only a small fraction of workers (4\%) earn more than \$7.25 / hour \cite{hara2018data}.

Recent research has also identified how crowdworking platforms often necessitate various kinds of unpaid labor from crowdworkers, which reduces overall wages.  

For example, one report found that for every hour of paid work, workers spend another 18 minutes on unpaid work, including searching for tasks \citep{berg2015income}. 
Another recent study found that once daily invisible labor was accounted for, the median hourly wage for crowdworkers on Amazon Mechanical Turk dropped from \$3.76 to \$2.83 \citep{Toxtli2021}.

Workers often invest significant labor outside the platform itself to find tasks, relying on web browsers extensions and participating in crowd work forums \cite{hanrahan2021expertise, kaplan2018striving}. Time spent working is compounded by competition from other crowdworkers \citep{semuels2018internet}, which can pressure workers to be constantly available to look for work \citep{berg2015income}. 

The working conditions of crowdworkers are characterized by long working hours, partially as a result of this competition. As \citet{berg2015income} notes, this conflicts with the work flexibility motivates many workers to choose crowd work.

Worker psychological safety is a particular area of concern. Crowd workers who work on content moderation of user generated content often need to look at content that includes violent imagery or sexual and pornographic content \cite{roberts2016digital}, or to transcribe conversations about trafficking children into sexual slavery \cite{emerson2019suffering}. 
In many cases, it is impossible to ascertain that a job may contain such content \cite{emerson2019suffering}.
If crowdworkers find themselves upset or disturbed by this content, they have little recourse; often, workers need to sign non-disclosure agreements preventing them from talking to anyone about the awful things they must look at, even for support \cite{roberts2016digital}. Additionally, raising concerns to their employers is quite difficult; both bureaucracy and physical distance (many of these workers are in the Global South) prohibit any direct lines of feedback or complaints. There is research available on the long-term impacts of viewing harmful user-generated content, but it is difficult to assess the full harm this causes to workers' well-being \cite{roberts2016digital}.

\subsection{Power dynamics}
Power dynamics between the requesters and annotators is another major challenge. Annotators are often heavily distanced from those leading the development of datasets are requesting tasks, which can obfuscate working conditions.
Top-down organizational structures often results in the workers viewing requesters as more informed as they are the ones who provided the data and the label schema \citep{miceli2020between}. Hence, instead of resolving ambiguities, workers are more likely to try to judge from the standpoint of the requester, often with limited exposure to the goals of the annotation. This contributes to the \textit{portability trap} \citep{selbst2018fairness}: a ``failure to understand how repurposing algorithmic solutions designed for one social context may be misleading, inaccurate, or otherwise do harm when applied to a different context.''

Power dynamics are also at play in the rejection of work: a large majority of crowdworkers (94\% as per \citep{berg2015income}) have had work that was rejected or for which they were not paid. Yet, some platforms give requesters full rights over the data they receive, regardless of whether they accept or reject it, and workers have no way of taking legal action if requesters use rejected work anyway \cite{irani2013turkopticon}; \citet{roberts2016digital} describes this system as one that ``enables wage theft''. Moreover, rejecting work and withholding pay is painful because rejections are often caused by unclear instructions and the lack of meaningful feedback channels. Many crowdworkers report that poor communication negatively affects their work \citep{berg2015income}. Moreover, requesters get to choose whether the work is up to their standards before choosing whether to pay for it, even though rejections are often caused by unclear instructions and the very lack of feedback channels they refuse to provide \cite{berg2015income}. Workers also feel powerless to speak up about perceived injustices from requesters or the platform; Amazon Mechanical Turk (AMT) users have reportedly had their accounts suspended for speaking negatively about Amazon \cite{semuels2018internet}. Additionally, requesters can block users who offer them feedback without consequence \cite{berg2015income}.

Power asymmetries also reflect global power dynamics. For instance, since technology development happens primarily in the West, human computation from the Global South is often relegated to the margins \citep{sambasivan2021re}. In particular, \cite{sambasivan2021re} points out that the technical, social, ethical, and physical distance between the  builders of a technology and the communities it is meant to serve is large, in such settings. 
\cite{bott2012role} has pointed out the potential of crowdsourcing to revolutionize civic participation in many developing countries to address complex challenges in governance around global issues such as climate change, poverty, armed conflict, and other crises. They also point out the challenges when it comes to employing crowdsourced interventions on the ground in the Global South. They note that systemic disparities endemic to local contexts are often reflected in who is represented in \textit{crowd}; for instance, the digital crowd in the Global South tends to over-represent the elite, educated, young males who belong to the upper tiers of local social hierarchies.

\citet{roberts2016digital} compares the commercial content moderation work to the practice of developed nations offloading their hazardous e-waste refuse on countries in the Global South. Her interviewees characterized this work as ``akin to being immersed in `a cesspool' -- feeling that they are within a pit of toxic matter and waste day in and day out''. The metaphor goes further in highlighting the fact that digital content moderation when outsourced to countries in the Global South, serves to keep the digital refuse away from the field of vision of those in the Global North who are responsible for its existence, and for whom it was intended, in much the same way the rotting garbage and e-waste produced in the Global North is kept away.

On the other hand, some platforms have geographical blocking, which many non-Americans find problematic \cite{berg2015income} since it can be used to exclude them. This reinforces the dynamic where requesters in the United States get to decide which global perspectives they want to consider for their task, and which they want to disregard.

The anonymous and geographically distributed nature of crowdsourced annotation work imposes significant barriers to collective action on the part of dataset annotators. While some platforms offer communication spaces, such as discussion forums, for workers to communicate with one another, these platform-moderated spaces have been shown to be ineffective at supporting labor organizing or worker power \cite{Gerber21}. In response, several tools have been developed independently from crowdwork platforms to support crowd workers.

For example, TurkerNation, Turk Alert, MTurkGrind, and Reddit’s /r/HITsWorthTurkingFor offer online forums for AMT workers to share information about well-paying work and share experiences with different requesters and Turkopticon \cite{irani2013turkopticon, turkopticon} is a browser add-on that enables AMT workers to review and report requesters and view reviews from other workers. These tools can help workers overcome  the information asymmetries built into the AMT platform \cite{Martin2014}. 
Dynamo is another community platform designed specifically to support and enable collective action for AMT workers, creating ``unities without unions'' \cite{Salehi2015}. A 2015 study of the platform found that twenty-two ideas for action had been generated and two active campaigns had been initiated.

In summary, responsible data annotation requires careful consideration of the power dynamics that structure the working relationship between requesters, annotators, and the platforms. 

\cite{semuels2018internet} As a result, workers feel pressured to be constantly available online to look for work and work longer hours \cite{berg2015income}. \cite{berg2015income} notes that this is in conflict with one of the reasons many workers choose crowdwork, which is flexibility in working hours.

\section{CrowdWorkSheets: A Documentation Framework for Crowdsourced Dataset Annotation}
\label{framework}
We now introduce our framework, CrowdWorkSheets, which outlines a series of \textit{considerations} designed to guide the collection, use, and dissemination of crowd-sourced annotations and \textit{questions} designed to elicit information about various decisions and outcomes. We have decomposed the framework into sections based on different parts of a typical dataset construction pipeline, from the formulation of tasks to dissemination of datasets.

\subsection{Task formulation}
First, we must ask: \textit{what are we asking annotators to do?} 
Our considerations and documentation questions focus on many aspects of task formulation including which assumptions we make about annotators, how we handle ambiguity and subjectivity within our task, and how our task is ultimately framed and communicated. 

While some tasks tend to pose objective questions with a correct answer (\textit{is there a human face in an image?}), oftentimes datasets aim to capture judgement on relatively subjective tasks with no universally correct answer (\textit{is this piece of text offensive?}). Moreover, even seemingly objective tasks can still be rife with ambiguity or corner-cases and ultimately require subjective judgements to be made on the part of annotators. As such, it is important to consider how questions afford varied interpretations or may require subjective judgements on the part of the annotators.  
Clarifying such aspects of of an annotation task as critical to ensuring a resulting dataset captures the aspects of human intelligence they are meant to capture. Moreover, as discussed in Section~\ref{experiences}, a survey of crowdworkers on AMT found that many instances of work rejection were due to unclear instructions \citep{berg2015income}. 

While we discuss the nuances of annotator selection in greater depth in Section~\ref{selecting_annotators}, tasks should be formulated based on considerations regarding \textit{who} will be annotating data and what perspectives should (or should not) be included. Determinations should be tied to the purpose of dataset creation and the downstream use cases it is meant to serve, rather than what is convenient, efficient, or scalable. 
Some tasks may benefit from being informed by the annotators' lived experiences and thus may be designed to explicitly seek out such expertise. On the other hand, a dataset developer may want to frame task instructions so as to  restrict the annotators from relying on their lived experiences, e.g. for a dataset meant to capture a set of policies defined by a platform. 

Finally, when formulating a task, it is important to consider how much information to disclose to annotators about the task in advance. Some information may be essential to disclose in order to enable annotators to make informed decision regarding whether or not to accept the task. For example, disclosure of how data will be stored, packaged, and potentially published may be particularly important when sociodemographic, or other sensitive information, about annotators is being requested. Similarly, disclosure of risks relating to psychological harm should be included where appropriate.

\textit{Considerations}
\begin{itemize}[itemsep=1pt,topsep=0pt]
    
    \item Consider the role subjectivity plays in your annotation task. Remember that individuals with different social and cultural backgrounds might differ in their judgements.
    
    \item Consider the forms of expertise that should be incorporated through data annotation, including both formal disciplinary training and lived experience with the problem domain. Remember that insufficiently capturing this expertise in the annotator pool may carry risks for downstream model usage.
    \item 
    Make sure task instructions are clear and unambiguous in order to prevent annotators from wasting time on a task where their work will be rejected due to misunderstandings. Consider assessing the task instructions in a small-scale setting prior to launching your full annotation task.
    
    \item Consider the personal information you are collecting from annotators and the potential ethical or privacy risks that may accompany such collection.
    
    \item Consider the amount of information you disclose to annotators prior to engagement with the task and ensure annotators have an opportunity to make informed decisions based on any potential risks the task carries. 
\end{itemize}

\textit{Documentation questions}
\begin{enumerate}[itemsep=1pt,topsep=0pt]
    \item At a high level, what are the subjective aspects of your task?
    
    \item What assumptions do you make about annotators?
    
    \item How did you choose the specific wording of your task instructions? What steps, if any, were taken to verify the clarity of task instructions and wording for annotators?
    
    \item What, if any, risks did your task pose for annotators and were they informed of the risks prior to engagement with the task?
    
    \item What are the precise instructions that were provided to annotators?
\end{enumerate}

\subsection{Selecting annotators}
\label{selecting_annotators}
Next, we ask: \textit{who is annotating the data?} While there is no single ``correct'' way to assemble an annotator pool, the selection of an annotator pool is a highly consequential decision. Since annotators from different communities can produce significantly different annotations given the same task \cite{sen2015turkers}, it is important to recognize that annotator selection may have a significant impact on the labels of your dataset. With this in mind, it is important to consider the intended use of the datasets---which communities will be most impacted by models built from the data, and which communities could be harmed the most by resulting biases present if they are not represented in the annotator pool?

In some cases, social identities of annotators indicate a form of expertise relevant to our task so it may be prudent to select annotators based on self-identified sociodemographic factors. In other cases, it may be important to select annotators based on other forms of expertise or experience with a problem domain. Understanding one's desired annotator pool may subsequently impact decisions regarding platform selection, as different platforms offer differing degrees of flexibility to assemble custom annotator pools. 

While selecting annotators based on sociodemographic factors may help ensure a dataset reflects perspectives of certain groups, targeted data collection efforts---particularly those oriented towards the inclusion of marginalized groups---are not without risk. For example, \citep{Denton2020BringingTP} discuss how the mere inclusion of marginalized groups within a dataset, without sufficient attention to broader considerations of data capture and use, can operate as a form of ``predatory inclusion''\footnote{The term "predatory inclusion" has been used to describes modes of inclusion that are extractive and predatory in nature in other domains (e.g. \cite{seamster2017})}. Discourses of inclusion can serve to ``further rather than subvert vulnerability to what might more broadly be called `data violence'" \citep{termsofinclusion}. 
From a privacy perspective, if sociodemograph-ic information is collected and published with a dataset, developers should take extra care to mitigate risks of unintentionally making annotators identifiable. 

\textit{Considerations}
\begin{itemize}[itemsep=1pt,topsep=0pt]
    \item While there are multiple valid ways to assemble an annotator pool,  remember that annotators are not interchangeable, and that 
    the decisions in this stage can heavily impact the final dataset. 
    \item Consider the ways in which social identities of annotators may relate to the forms of expertise important for the task.
    \item Consider the intended usage contexts of the dataset, and the marginalized communities therein, when choosing which annotators to be prioritized to be included. 
    \item Consider how labor practices intersect with the choice of who the annotators are. For example: if female annotators make up the majority, as they do in the U.S.\ \citep{posch2018characterizing}, consider how fair payment, or a lack thereof, could impact this group.
\end{itemize}

\textit{Documentation questions}
\begin{enumerate}[itemsep=1pt,topsep=0pt]
    \item Are there certain perspectives that should be privileged? If so, how did you seek these perspectives out?
    
    \item Are there certain perspectives that would be harmful to include? If so, how did you screen these perspectives out?

    \item Were sociodemographic characteristics used to select annotators for your task? If so, please detail the process.
    
    \item If you have any aggregated sociodemographic statistics about your annotator pool, please describe.
    
    \item Do you have reason to believe that sociodemographic characteristics of annotators may have impacted how they annotated the data? Why or why not?
    
    \item Consider the intended context of use of the dataset and the individuals and communities that may be impacted by a model trained on this dataset. Are these communities represented in your annotator pool?
\end{enumerate}

\subsection{Platform and infrastructure choices}
Next, we ask, \textit{under what conditions are data annotated?} As described in Section~\ref{experiences}, platform policies around compensation and power asymmetries play a huge role in shaping worker experiences and the quality of work that annotators produce. Different platforms offer different affordances for communication between task requesters and annotators, which might impact the extent to which task requesters can incorporate annotator feedback into the task framing or annotator guidelines. 
Different platforms also impose different minimum-pay constraints; requesters may want to support platforms that uphold fair pay standards. 
Additionally, requesters should be mindful of potential differences between legal minimum wages and a living wage \cite{living-wage}.

Separately from the platform, task creators should be aware of worker pay per hour; some platforms may only offer requesters the option to select pay per item for an annotation task, and the defaults may be set low. Task creators should take care when estimating work time per item to ensure they are paying workers fairly.
Another thing to consider when choosing a platform for data annotation is how well that platform supports rater psychological safety. Some platforms provide more affordances than others for crowdworkers to seek out support if they are experiencing distress, or if they otherwise have questions or feedback for requesters.

\textit{Considerations}
\begin{itemize}[itemsep=1pt,topsep=0pt]
    \item Consider platform's underlying annotator pool and the options they provide to source specialized rater pools, and whe-ther they enable you to curate an appropriate pool of annotators (e.g.\ considering sociodemographic factors or domain expertise). 
    \item Consider comparing and contrasting the minimum pay requirements established across different platforms. You may choose to support a platform that upholds fair pay standards.  
    \item Consider the extent to which you would like to establish a channel of communication and feedback between your team and the annotators. Platform mediated channels of communication can give annotators an opportunity to provide feedback on confusing instructions, or otherwise seek out support.
\end{itemize}

\textit{Documentation questions}
\begin{enumerate}[itemsep=1pt,topsep=0pt]
    \item What annotation platform did you utilize?

    \item At a high level, what considerations informed your decision to choose this platform?
    
    \item Did the chosen platform sufficiently meet the requirements you outlined for annotator pools? Are any aspects not covered?
    
    \item What, if any, communication channels did your chosen platform offer to facilitate communication with annotators? How did this channel of communication influence the annotation process and/or resulting annotations?
    
    \item How much were annotators compensated? Did you consider any particular pay standards, when determining their compensation? If so, please describe.
\end{enumerate}

\subsection{Dataset analysis and evaluation}
Once data instances are annotated, \textit{what do we do with the results?} This section focuses on considerations related to the process of converting the ``raw'' annotations into the labels that are ultimately packaged in a dataset. A common practice in building crowdsourced annotations for discrete labeling tasks is to obtain multiple annotator judgements that are then aggregated (e.g.,\ through majority voting) to obtain a single ``ground truth'' that is released in the dataset \citep{sabou2014corpus}. However, the disagreements between annotators may embed valuable nuances about the task \citep{alm2011subjective,aroyo2013crowd}. Aggregation, in such cases may obscure such nuances, and potentially exclude perspectives from minority annotators \citep{prabhakaran2021releasing}.
It is thus critical to consider uncertainty and disagreement between annotators, and potentially leverage this as a signal, to avoid losing nuanced and diverse opinions in the aggregation process. It might be important to analyze how annotators disagree along sociodemographic lines in order to be able to share this information with potential users of the dataset, so they can best understand how to represent these diverse perspectives in their use of the data.

\textit{Considerations}
\begin{itemize}[itemsep=1pt,topsep=0pt]
    \item Consider including uncertainty or disagreement between annotations on each instance as a signal in the dataset.
    \item Consider analyzing systematic disagreements between annotators of different sociodemographic groups in order to better understand how diverse perspectives are represented.
    \item Consider how the final dataset annotations will relate to individual annotator responses. For instance, one option is to release only the aggregated labels, e.g.\ through a majority vote.
    Consider what valuable information might be lost through such aggregation.
\end{itemize}

\textit{Documentation questions}
\begin{enumerate}[itemsep=1pt,topsep=0pt]
    \item How do you define the annotation quality in your context, and how did you assess quality in your dataset?

    \item Have you conducted any analysis on disagreement patterns? If so, what analyses did you use and what were the major findings?
    
    \item Did you analyze potential sources of disagreement? 
    
    \item How do the individual annotator responses relate to the final labels released in the dataset? 
\end{enumerate}

\subsection{Dataset release and maintenance}
Finally, it is critical to consider \textit{what is the future of the dataset?} Data exists within an ever-changing world, and should be viewed and used in that context. Users of the dataset now and in the future should understand the limitations of the data based on when and how it was collected. For example, a dataset may require periodic updates to remain robust to new slang or changes in language use over time. In addition, annotation tasks may be predicated upon legal definitions or medical standards that may change according to decisions by institutions or governing bodies.

\textit{Considerations}
\begin{itemize}[itemsep=1pt,topsep=0pt]
\item Consider designing and sharing a dataset maintenance plan \citep{Hutchinson2021}.
\item Consider potential conditions under which annotations may become outdated or less useful.
\end{itemize}

\textit{Documentation Questions}:
\begin{enumerate}[itemsep=1pt,topsep=0pt]
    \item Do you have reason to believe the annotations in this dataset may change over time? Do you plan to update your dataset?
    
    \item Are there any conditions or definitions that, if changed, could impact the utility of your dataset?
    
    \item Will you attempt to track, impose limitations on, or otherwise influence how your dataset is used? If so, how?
    
    \item Were annotators informed about how the data is externalized? If changes to the dataset  are made, will they be informed?
    
    \item Is there a process by which annotators can later choose to withdraw their data from the dataset? Please detail.
    
\end{enumerate}

\section{Case Study}
We now present a \textit{hypothetical} case study to demonstrate how our considerations outlined in Section ~\ref{framework} might be incorporated in practice and how dataset annotation decisions might be documented using CrowdWorkSheets. Responses to documentation questions are not intended to be prescriptive, nor are they completely comprehensive. Instead, they should be considered as one of many valid responses to this line of inquiry, and as a way to provoke further thought and discussion.

In this hypothetical case study, we take our goal to be the development of a benchmark dataset for public release to support academic research in social media content moderation. A Twitter corpus of 20,000 English-language tweets has been collected and we seek to label each tweet independently on a four-point ``toxicity'' scale defined in \cite{dixon2018annotation}.
\vspace{4mm}

{
\centering

\begin{tabular}{m{.97\columnwidth}c}
\hline
\multicolumn{1}{|c|}{\textbf{Task Formulation}} \\
\hline
\\
\end{tabular}
}

{
\small
    \textbf{At a high level, what are the subjective aspects of your task?} \\
    Judgements of toxicity of online comments is highly subjective. What makes a tweet harmful or hurtful varies greatly not only by the literal content of the tweet, but by the context surrounding it. In our task setup, tweets are presented to annotators in isolation, so they do not have access to the overall context of the online conversation. As such, we anticipate that annotators may infer surrounding context and make subjective judgements based on this inference. \\
    
    \textbf{What assumptions do you make about annotators?} \\
    Some of the key assumptions we make of our annotators:
    \begin{itemize}[after=\vspace{\baselineskip}]
        \item Annotators that claim proficiency in English and familiarity with social media have enough context to reasonably interpret the task.
        \item By giving a clear understanding of the goals of this work and explicitly indicating that this is a subjective task where disagreement is expected, we will increase the likelihood that annotators will allow their lived experiences to inform how they label toxicity.
        \item By paying well, we will increase the likelihood that annotators will take time to think through particularly challenging examples.
    \end{itemize} 
    
    \textbf{How did you choose the specific wording of your task instructions? What steps, if any, were taken to verify the clarity of task instructions and wording for annotators?} \\
    To align with existing research in the area, we've chosen to give annotators an existing definition of toxicity, as ``rude, disrespectful or otherwise likely to make someone leave a discussion'' \cite{aroyo2019crowdsourcing}.
    To settle on a final task wording, our research team first completed 50 annotation tasks each to identify any obvious challenges applying this definition. We then ran several small pilot studies with slightly varying task instructions, and allowed annotators the option to give feedback on aspects that were unclear. Looking over these results, we settled on the question phrasing that yielded the least reported confusion.
    We intentionally chose to leave our definition of toxicity somewhat open to interpretation, operating under the understanding that being overly specific in task instructions for subjective work does not improve response quality \cite{aroyo2015truth}. We also explicitly informed annotators that we expect a variety of interpretations of each comment, and that we were looking for their personal best judgements in the given situation. 
    To motivate thoughtful responses, we chose to pay well above minimum page and gave annotators a clear idea of the ultimate purpose of their work.
    However, we know that it is inevitable that some annotators will simply give answers they think we want as quickly as possible. While we screen out responses below a minimum duration, it's impossible to ensure every answer is honest and thoughtful. We assume that these responses are randomly distributed; we leave it up to dataset users to do further analysis.\\
    
    \textbf{What, if any, risks did your task pose for annotators and were they informed of the risks prior to engagement with the task?}\\
    Our task required annotators to read text that potentially contained hate speech, slurs, and other harmful content. As such, the task posed a risk of psychological harm to annotators. Moreover, given that we selected annotators who had previously experienced online harassment, there is a potential for the task to trigger an emotional response related to past trauma. We informed annotators about this risk prior to the start of the task. We also informed annotators that we would be requesting sociodemographic information in order to assess disagreement across different groups. We outlined our data storage policy and steps we took to prevent responses from being linked to sociodemographic information. \\
    
    \textbf{What are the precise instructions that were provided to annotators?}\\
    The final task instructions used for data collection reflected in the released data is available at HypotheticalTaskInstructions.com. 
}

\vspace{4mm}

{
\centering

\begin{tabular}{m{.97\columnwidth}c}
\hline
\multicolumn{1}{|c|}{\textbf{Selecting Annotations}} \\
\hline
\\
\end{tabular}
}

{
\small
    \textbf{Are there certain perspectives that should be privileged? If so, how did you seek these perspectives out?}\\
        We want to privilege the perspectives of annotators who have personally experienced online harassment or hold marginalized identities that are often targeted online. To this end, we included screening questions such that our annotator pool consisted of raters who have direct experience with online harassment. We intentionally defined ``direct experience'' very broadly to capture a wide range of experiences, intending to include annotators who've been personally harassed by others via online channels, who've encountered online content that threatened or disparaged identities they share, who have experience moderating online forums, or who have felt otherwise personally affected by harmful online content.\\
        
        
    \textbf{Are there certain perspectives that would be harmful to include? If so, how did you screen these perspectives out?}\\
        We believe that there are many harmful worldviews annotators might hold that we do not want captured by our annotations; we do not want to employ annotators who participate in hateful online communities, for example. To attempt to account for this, we identified several tweets that we agreed were unambiguously toxic, and screened out any annotators that did not label these as toxic.\\
        
    \textbf{Were sociodemographic characteristics used to select annotators for your task? If so, please detail the process.}\\
        In addition to screening for annotators who have previously experienced online harassment, we selected annotators based on self-identified gender and age. We aimed for an approximately gender balanced pool and we selected for at least 10\% of the annotators to be older than 65 years old. Because annotators were sourced from multiple geographic regions, we could not easily specify thresholds for racial or ethnic diversity; however, because we are screening for annotators who have experienced harassment online, we achieved decent representation among marginalized groups.\\
        
    \textbf{If you have any aggregated sociodemographic statistics about your annotator pool, please describe.}\\
        We first selected annotators who indicated that they had previously experienced online harassment. This resulted in a pool that is disproportionately composed of women and people of color compared with platform demographics. More specific demographic breakdowns are available with the released dataset.\\
        
    \textbf{Do you have reason to believe that sociodemographic characteristics of annotators may have impacted how they annotated the data? Why or why not?}\\
        Yes, we believe that annotators who have themselves experienced online harassment may be more likely to identify tweets as toxic. Based on rates of reported experience with hate speech attacks, we also expect that these annotators will disproportionately be members of marginalized social groups in their respective geographic region. \\
        
    \textbf{Consider the intended context of use of the dataset and the individuals and communities that may be impacted by a model trained on this dataset. Are these communities represented in your annotator pool?}\\
        Our intended audience is researchers studying English-language online content moderation, although we can anticipate that our work may have impact within industry. Content moderation has far-reaching and pervasive influence on online discourse, which impacts a wide range of individuals and communities.  Not everyone is equally vulnerable to the worst impacts of toxic language online, so we specifically selected for an annotator pool where this more vulnerable population is represented.\\
}

\vspace{4mm}

{
\centering

\begin{tabular}{m{.97\columnwidth}c}
\hline
\multicolumn{1}{|c|}{\textbf{Platform and Infrastructure Choices}} \\
\hline
\\
\end{tabular}
}

{
  \small
  \textbf{What annotation platform did you utilize?}\\
    We're using HypotheticalPlatform.\\

    \textbf{At a high level, what considerations informed your decision to choose this platform?}\\
        We have selected HypotheticalPlatform for several reasons: First, they are a generally reliable platform with a history of high data quality. Second, they are able to guarantee that annotators are paid at or above a living wage. Third, their platform's interface allows annotators to easily communicate feedback and concerns. And finally, their platform allows us to  make ample use of screening questions to select the annotator pool for our main body of work.\\
    
    \textbf{Did the chosen platform sufficiently meet the requirements you outlined for annotator pools? Are any aspects not covered?}\\
        We were able to meet all of our requirements for annotator pools through the use of many screening and demographic questions. The main trade-off we made to accomplish this is in cost; to pay annotators well, including for their time answering screening questions, we set a limit on the number of tweets we could label.\\
    
    \textbf{What, if any, communication channels did your chosen platform offer to facilitate communication with annotators? How did this channel of communication influence the annotation process and/or resulting annotations?} \\
        We included a free response section at the end of our survey to allow feedback from annotators. In our pilot studies, we used this to clarify our task instructions. In the full study, most annotators left this blank, so we chose to leave them out of the final dataset.\\
    
    \textbf{How much were annotators compensated? Did you consider any particular pay standards, when determining their compensation? If so, please describe.} \\
        Informed by the 2020 results of the MIT Living Wage Calculator \cite{living-wage}, we aimed for annotators to take home at least \$25/hr on our work, with the goal of comfortably reaching a living wage for a single adult with no dependents, and decrease the pressure to complete tasks as quickly as possible. Annotators were paid \$6.25 for labeling a batch of 40 tweets, designed to take no more than 15 minutes, and verified over the course of the annotation job.
}

\vspace{4mm}

{
\centering

\begin{tabular}{m{.97\columnwidth}c}
\hline
\multicolumn{1}{|c|}{\textbf{Dataset Analysis and Evaluation}} \\
\hline
\\
\end{tabular}
}

{
\small
 \textbf{How do you define the quality of annotations in your context, and how did you assess the quality in the dataset you constructed?}\\
    We assessed quality along several dimensions, each of which had an associated question in each 40-question batch:
    \begin{itemize}
        \item Attention: We included 1 attention check question was introduced that instructs the annotator to give a particular response so ensure annotators are reading each question; 
        \item Self-consistency:  We included 2 duplicated questions within each batch, to ensure annotators were actually reading each tweet and being self-consistent in their responses. 
        \item Alignment with pre-defined ratings: We included several 2 tweets that the research team had pre-labeled as unoffensive and highly offensive. We chose tweets for which we would expect no disagreement from annotators.
    \end{itemize}
    We removed from the final dataset all batches where 2 or more of these 5 data quality questions were incorrectly answered. This ultimately accounted for 12\% of our data.\\
    
   \textbf{Have you conducted any analysis on disagreement patterns? If so, what analyses did you use and what were the major findings?}\\
        While the main purpose of this work is data collection and not analysis, we did conduct very preliminary analyses as a starting point for dataset users.  We ran standard inter-annotator agreement metrics and found a relatively low interannotator agreement across all raters (Fleiss' $\kappa = 0.25$ \cite{fleiss1971measuring}). However, we do not believe this to be an issue of data quality---when we looked at the data aggregated along different demographic axes, we found many demographic groups with high interannotator agreement whose annotations differ significantly from the majority opinion.\\

    \textbf{Did you analyze potential sources of disagreement?} \\
        In our preliminary analysis, we looked at a few annotator demographics as a source of disagreement. There are a myriad of other factors one could analyze with respect to disagreement---tweet topic, presence or absence of particular words, or how quickly annotators responded, for example---but as this is intended to be released as a research dataset, we have not conducted all of these analyses.\\
    
    \textbf{How do the individual annotator responses relate to the final labels released in the dataset?}\\
        After bucketing annotator demographics such that no annotator was uniquely identifiable, we released all responses, attached to the demographics of the annotator that gave each response. We chose not to aggregate responses into final tweet toxicity labels, and instead leave this to dataset users to aggregate in a way that's appropriate for their use case.
    
}
\vspace{4mm}

{
\centering

\begin{tabular}{m{.97\columnwidth}c}
\hline
\multicolumn{1}{|c|}{\textbf{Dataset Release and Maintenance}} \\
\hline
\\
\end{tabular}
}

{
\small
    \textbf{Do you have reason to believe the annotations in this dataset may change over time? Do you plan to update your dataset?}\\
        The relevancy of and perceptions about tweets will certainly change over time. In an effort to remind dataset users that this data should be taken in its temporal context, we include the month and year that each tweet was (a) written and (b) annotated as meta-data.  However, as a longer-term strategy, we are also open-sourcing and making public all parts of our annotation pipeline, including rater instructions, data formatting schemes, and information on how to coordinate with our data labeling partners. We will publicly extend an open invitation to future collaborators who want to reuse our pipeline to annotate more data. If this pipeline is used and our guidelines followed satisfactorily, we will append future annotations to our existing dataset.\\
    
    \textbf{Are there any conditions or definitions that, if changed, could impact the utility of your dataset?}\\
        Over time we expect societal views to deviate somewhat from the annotations collected. For example, it will not capture any shifts in attitude regarding language targeting social groups that may be considered marginalized in the future but that are not considered marginalized today. \\
    
    \textbf{Will you attempt to track, impose limitations on, or otherwise influence how your dataset is used? If so, how?}\\
        To access the data, we require dataset users indicate their affiliation, contact information, and use case. The research team will be assessing uses on a case-by-case basis, with particular attention given to risks associated with use cases that explicitly include sociodemographic data in their modeling. We also ask that any publications cite our dataset release paper so we can track academic uses of the dataset. Our full data license if available at HypotheticalDataLicense.com.\\
    
     \textbf{Were annotators informed about how the data is externalized? If changes to the dataset are made, will they be informed?}\\
        Annotators were informed that this data will be released as a research dataset prior to engaging in the task. We allowed raters to opt in to an email list that with share updates about data release an availability. This site will contain an automatically-updated list of papers that cite our dataset release paper. \\
    
    \textbf{Is there a process by which annotators can later choose to withdraw their data from the dataset? If so, please detail.}\\
        By design, we have no mechanisms of linking individual annotators to specific responses, and so have no option for annotators to withdraw their annotations form our dataset. We make this explicit to the annotators, and allow them to stop answering questions at any point if they decide they no longer want to continue. 
}

\section{Conclusion}
 In this work, we challenge the common portrayal of dataset annotators as interchangeable. Rather, we argue, their individual histories and experiences bring unique perspectives to the table that can become encoded in the overall dataset in a significant ways. Therefore, it becomes imperative to consider how the process of selecting annotators, and their experience working on annotation, is documented alongside other aspects of dataset development. Towards this end, we introduced CrowdWorkSheets, a framework for reflecting on and documenting key decision points of crowdsourced dataset development, and a set of recommendations for dataset developers. While this framework is oriented towards individual dataset developers, we also recognize the role large institutions can play in shifting incentives to engage with these recommendations, e.g.  incentivizing transparent  dataset documentation through conference submission and reviewer guidelines. 
\newline \newline
\textbf{Funding:} This research was supported by Google.

\hypersetup{
    breaklinks=true,   
    colorlinks=true,   
    }

\bibliographystyle{ACM-Reference-Format}
\bibliography{main}


\begin{thebibliography}{59}


\ifx \showCODEN    \undefined \def \showCODEN     #1{\unskip}     \fi
\ifx \showDOI      \undefined \def \showDOI       #1{#1}\fi
\ifx \showISBNx    \undefined \def \showISBNx     #1{\unskip}     \fi
\ifx \showISBNxiii \undefined \def \showISBNxiii  #1{\unskip}     \fi
\ifx \showISSN     \undefined \def \showISSN      #1{\unskip}     \fi
\ifx \showLCCN     \undefined \def \showLCCN      #1{\unskip}     \fi
\ifx \shownote     \undefined \def \shownote      #1{#1}          \fi
\ifx \showarticletitle \undefined \def \showarticletitle #1{#1}   \fi
\ifx \showURL      \undefined \def \showURL       {\relax}        \fi
\providecommand\bibfield[2]{#2}
\providecommand\bibinfo[2]{#2}
\providecommand\natexlab[1]{#1}
\providecommand\showeprint[2][]{arXiv:#2}

\bibitem[tur(2008)]%
        {turkopticon}
 \bibinfo{year}{2008}\natexlab{}.
\newblock \bibinfo{title}{Turkopticon}.
\newblock \bibinfo{howpublished}{\url{https://turkopticon.net/}}.
\newblock
\newblock
\shownote{Accessed: 2021-07-21}.


\bibitem[Alkhatib et~al\mbox{.}(2017)]%
        {Alkhatib2017}
\bibfield{author}{\bibinfo{person}{Ali Alkhatib}, \bibinfo{person}{Michael~S.
  Bernstein}, {and} \bibinfo{person}{Margaret Levi}.}
  \bibinfo{year}{2017}\natexlab{}.
\newblock \showarticletitle{Examining Crowd Work and Gig Work Through The
  Historical Lens of Piecework}. In \bibinfo{booktitle}{\emph{Proceedings of
  the 2017 CHI Conference on Human Factors in Computing Systems}} (Denver,
  Colorado, USA) \emph{(\bibinfo{series}{CHI '17})}.
  \bibinfo{publisher}{Association for Computing Machinery},
  \bibinfo{address}{New York, NY, USA}, \bibinfo{pages}{4599–4616}.
\newblock
\showISBNx{9781450346559}
\urldef\tempurl%
\url{https://doi.org/10.1145/3025453.3025974}
\showDOI{\tempurl}


\bibitem[Aroyo et~al\mbox{.}(2019)]%
        {aroyo2019crowdsourcing}
\bibfield{author}{\bibinfo{person}{Lora Aroyo}, \bibinfo{person}{Lucas Dixon},
  \bibinfo{person}{Nithum Thain}, \bibinfo{person}{Olivia Redfield}, {and}
  \bibinfo{person}{Rachel Rosen}.} \bibinfo{year}{2019}\natexlab{}.
\newblock \showarticletitle{Crowdsourcing Subjective Tasks: The Case Study of
  Understanding Toxicity in Online Discussions}. In
  \bibinfo{booktitle}{\emph{Companion Proceedings of The 2019 World Wide Web
  Conference}} (San Francisco, USA) \emph{(\bibinfo{series}{WWW '19})}.
  \bibinfo{publisher}{Association for Computing Machinery},
  \bibinfo{address}{New York, NY, USA}, \bibinfo{pages}{1100–1105}.
\newblock
\showISBNx{9781450366755}
\urldef\tempurl%
\url{https://doi.org/10.1145/3308560.3317083}
\showDOI{\tempurl}


\bibitem[Aroyo and Welty(2013)]%
        {aroyo2013crowd}
\bibfield{author}{\bibinfo{person}{Lora Aroyo} {and} \bibinfo{person}{Chris
  Welty}.} \bibinfo{year}{2013}\natexlab{}.
\newblock \showarticletitle{Crowd truth: Harnessing disagreement in
  crowdsourcing a relation extraction gold standard}.
\newblock \bibinfo{journal}{\emph{WebSci2013. ACM}} (\bibinfo{year}{2013}).
\newblock


\bibitem[Aroyo and Welty(2015)]%
        {aroyo2015truth}
\bibfield{author}{\bibinfo{person}{Lora Aroyo} {and} \bibinfo{person}{Chris
  Welty}.} \bibinfo{year}{2015}\natexlab{}.
\newblock \showarticletitle{Truth Is a Lie: Crowd Truth and the Seven Myths of
  Human Annotation}.
\newblock \bibinfo{journal}{\emph{AI Magazine}} \bibinfo{volume}{36},
  \bibinfo{number}{1} (\bibinfo{date}{Mar.} \bibinfo{year}{2015}),
  \bibinfo{pages}{15--24}.
\newblock
\urldef\tempurl%
\url{https://doi.org/10.1609/aimag.v36i1.2564}
\showDOI{\tempurl}


\bibitem[Bender and Friedman(2018)]%
        {bender2018data}
\bibfield{author}{\bibinfo{person}{Emily~M. Bender} {and}
  \bibinfo{person}{Batya Friedman}.} \bibinfo{year}{2018}\natexlab{}.
\newblock \showarticletitle{Data Statements for Natural Language Processing:
  Toward Mitigating System Bias and Enabling Better Science}.
\newblock \bibinfo{journal}{\emph{Transactions of the Association for
  Computational Linguistics}}  \bibinfo{volume}{6} (\bibinfo{year}{2018}),
  \bibinfo{pages}{587--604}.
\newblock
\urldef\tempurl%
\url{https://doi.org/10.1162/tacl_a_00041}
\showDOI{\tempurl}


\bibitem[Berg(2015)]%
        {berg2015income}
\bibfield{author}{\bibinfo{person}{Janine Berg}.}
  \bibinfo{year}{2015}\natexlab{}.
\newblock \showarticletitle{Income security in the on-demand economy: Findings
  and policy lessons from a survey of crowdworkers}.
\newblock \bibinfo{journal}{\emph{Comp. Lab. L. \& Pol'y J.}}
  \bibinfo{volume}{37} (\bibinfo{year}{2015}), \bibinfo{pages}{543}.
\newblock


\bibitem[Bott and Young(2012)]%
        {bott2012role}
\bibfield{author}{\bibinfo{person}{Maja Bott} {and} \bibinfo{person}{Gregor
  Young}.} \bibinfo{year}{2012}\natexlab{}.
\newblock \showarticletitle{The role of crowdsourcing for better governance in
  international development}.
\newblock \bibinfo{journal}{\emph{Praxis: The Fletcher Journal of Human
  Security}} \bibinfo{volume}{27}, \bibinfo{number}{1} (\bibinfo{year}{2012}),
  \bibinfo{pages}{47--70}.
\newblock


\bibitem[Chmielinski et~al\mbox{.}(2020)]%
        {chmielinski2020DNP}
\bibfield{author}{\bibinfo{person}{Kasia~S. Chmielinski},
  \bibinfo{person}{Sarah Newman}, \bibinfo{person}{Matt Taylor},
  \bibinfo{person}{Josh Joseph}, \bibinfo{person}{Kemi Thomas},
  \bibinfo{person}{Jessica Yurkofsky}, {and} \bibinfo{person}{Yue~Chelsea
  Qiu}.} \bibinfo{year}{2020}\natexlab{}.
\newblock \bibinfo{title}{The Dataset Nutrition Label (2nd Gen): Leveraging
  Context to Mitigate Harms in Artificial Intelligence}.
  (\bibinfo{year}{2020}).
\newblock
\urldef\tempurl%
\url{http://securedata.lol/}
\showURL{%
\tempurl}
\newblock
\shownote{NeurIPS Workshop on Dataset Curation and Security}.


\bibitem[Deng et~al\mbox{.}(2009)]%
        {imagenet_cvpr09}
\bibfield{author}{\bibinfo{person}{J. Deng}, \bibinfo{person}{W. Dong},
  \bibinfo{person}{R. Socher}, \bibinfo{person}{L.-J. Li}, \bibinfo{person}{K.
  Li}, {and} \bibinfo{person}{L. Fei-Fei}.} \bibinfo{year}{2009}\natexlab{}.
\newblock \showarticletitle{{ImageNet: A Large-Scale Hierarchical Image
  Database}}. In \bibinfo{booktitle}{\emph{CVPR09}}.
\newblock


\bibitem[Denton et~al\mbox{.}(2020)]%
        {Denton2020BringingTP}
\bibfield{author}{\bibinfo{person}{Emily~L. Denton}, \bibinfo{person}{A.
  Hanna}, \bibinfo{person}{Razvan Amironesei}, \bibinfo{person}{Andrew Smart},
  \bibinfo{person}{Hilary Nicole}, {and} \bibinfo{person}{M. Scheuerman}.}
  \bibinfo{year}{2020}\natexlab{}.
\newblock \showarticletitle{Bringing the People Back In: Contesting Benchmark
  Machine Learning Datasets}.
\newblock \bibinfo{journal}{\emph{ICML Workshop on Participatory Approaches to
  Machine Learning}} (\bibinfo{year}{2020}).
\newblock


\bibitem[Dixon(2018)]%
        {dixon2018annotation}
\bibfield{author}{\bibinfo{person}{Lucas Dixon}.}
  \bibinfo{year}{2018}\natexlab{}.
\newblock \bibinfo{title}{Annotation instructions for Toxicity with
  sub-attributes}.
\newblock
  \bibinfo{howpublished}{\href{https://github.com/conversationai/conversationai.github.io/blob/master/crowdsourcing_annotation_schemes/toxicity_with_subattributes.md}{https://github.com/conversationai/conversationai.github.io/}}.
\newblock
\newblock
\shownote{Accessed: 2021-01-19}.


\bibitem[Emerson(2019)]%
        {emerson2019suffering}
\bibfield{author}{\bibinfo{person}{Sarah Emerson}.}
  \bibinfo{year}{2019}\natexlab{}.
\newblock \showarticletitle{`I Can Hear the Suffering': Rev Exposes Freelance
  Transcribers to Violent, Disturbing Content}.
\newblock \bibinfo{journal}{\emph{Medium OneZero}} (\bibinfo{year}{2019}).
\newblock


\bibitem[Filippova et~al\mbox{.}(2019)]%
        {Filippova2019}
\bibfield{author}{\bibinfo{person}{Anna Filippova}, \bibinfo{person}{Connor
  Gilroy}, \bibinfo{person}{Ridhi Kashyap}, \bibinfo{person}{Antje Kirchner},
  \bibinfo{person}{Allison~C. Morgan}, \bibinfo{person}{Kivan Polimis},
  \bibinfo{person}{Adaner Usmani}, {and} \bibinfo{person}{Tong Wang}.}
  \bibinfo{year}{2019}\natexlab{}.
\newblock \showarticletitle{Humans in the Loop: Incorporating Expert and
  Crowd-Sourced Knowledge for Predictions Using Survey Data}.
\newblock \bibinfo{journal}{\emph{Socius}}  \bibinfo{volume}{5}
  (\bibinfo{year}{2019}), \bibinfo{pages}{2378023118820157}.
\newblock


\bibitem[Fleiss(1971)]%
        {fleiss1971measuring}
\bibfield{author}{\bibinfo{person}{Joseph~L Fleiss}.}
  \bibinfo{year}{1971}\natexlab{}.
\newblock \showarticletitle{Measuring nominal scale agreement among many
  raters}.
\newblock \bibinfo{journal}{\emph{Psychological bulletin}}
  \bibinfo{volume}{76}, \bibinfo{number}{5} (\bibinfo{year}{1971}),
  \bibinfo{pages}{378}.
\newblock


\bibitem[Freeman(2018)]%
        {freeman2018}
\bibfield{author}{\bibinfo{person}{Joshua~Benjamin Freeman}.}
  \bibinfo{year}{2018}\natexlab{}.
\newblock \bibinfo{booktitle}{\emph{Behemoth : a history of the factory and the
  making of the modern world}}.
\newblock \bibinfo{publisher}{W.W. Norton \& Company, Inc.,}.
\newblock


\bibitem[Gebru et~al\mbox{.}(2018)]%
        {Gebru2018}
\bibfield{author}{\bibinfo{person}{Timnit Gebru}, \bibinfo{person}{Jamie
  Morgenstern}, \bibinfo{person}{Briana Vecchione},
  \bibinfo{person}{Jennifer~Wortman Vaughan}, \bibinfo{person}{Hanna Wallach},
  \bibinfo{person}{Hal Daum{\'e}~III}, {and} \bibinfo{person}{Kate Crawford}.}
  \bibinfo{year}{2018}\natexlab{}.
\newblock \showarticletitle{Datasheets for datasets}.
\newblock \bibinfo{journal}{\emph{arXiv preprint arXiv:1803.09010}}
  (\bibinfo{year}{2018}).
\newblock


\bibitem[Geiger et~al\mbox{.}(2020)]%
        {geiger2020}
\bibfield{author}{\bibinfo{person}{R.~Stuart Geiger}, \bibinfo{person}{Kevin
  Yu}, \bibinfo{person}{Yanlai Yang}, \bibinfo{person}{Mindy Dai},
  \bibinfo{person}{Jie Qiu}, \bibinfo{person}{Rebekah Tang}, {and}
  \bibinfo{person}{Jenny Huang}.} \bibinfo{year}{2020}\natexlab{}.
\newblock \showarticletitle{Garbage in, Garbage out? Do Machine Learning
  Application Papers in Social Computing Report Where Human-Labeled Training
  Data Comes From?}. In \bibinfo{booktitle}{\emph{Proceedings of the 2020
  Conference on Fairness, Accountability, and Transparency}} (Barcelona, Spain)
  \emph{(\bibinfo{series}{FAT* '20})}. \bibinfo{publisher}{Association for
  Computing Machinery}, \bibinfo{address}{New York, NY, USA},
  \bibinfo{pages}{325–336}.
\newblock
\showISBNx{9781450369367}
\urldef\tempurl%
\url{https://doi.org/10.1145/3351095.3372862}
\showDOI{\tempurl}


\bibitem[Gerber(2021)]%
        {Gerber21}
\bibfield{author}{\bibinfo{person}{Christine Gerber}.}
  \bibinfo{year}{2021}\natexlab{}.
\newblock \showarticletitle{Community building on crowdwork platforms: Autonomy
  and control of online workers?}
\newblock \bibinfo{journal}{\emph{Competition \& Change}} \bibinfo{volume}{25},
  \bibinfo{number}{2} (\bibinfo{year}{2021}), \bibinfo{pages}{190--211}.
\newblock


\bibitem[Ghosh et~al\mbox{.}(2021)]%
        {ghosh-etal-2021-detecting}
\bibfield{author}{\bibinfo{person}{Sayan Ghosh}, \bibinfo{person}{Dylan Baker},
  \bibinfo{person}{David Jurgens}, {and} \bibinfo{person}{Vinodkumar
  Prabhakaran}.} \bibinfo{year}{2021}\natexlab{}.
\newblock \showarticletitle{Detecting Cross-Geographic Biases in Toxicity
  Modeling on Social Media}. In \bibinfo{booktitle}{\emph{Proceedings of the
  Seventh Workshop on Noisy User-generated Text (W-NUT 2021)}}.
  \bibinfo{publisher}{Association for Computational Linguistics},
  \bibinfo{address}{Online}, \bibinfo{pages}{313--328}.
\newblock
\urldef\tempurl%
\url{https://aclanthology.org/2021.wnut-1.35}
\showURL{%
\tempurl}


\bibitem[Glasmeier(2020)]%
        {living-wage}
\bibfield{author}{\bibinfo{person}{Amy~K Glasmeier}.}
  \bibinfo{year}{2020}\natexlab{}.
\newblock \showarticletitle{Living Wage Calculator}.
\newblock  (\bibinfo{year}{2020}).
\newblock
\urldef\tempurl%
\url{livingwage.mit.edu}
\showURL{%
\tempurl}


\bibitem[Gray and Suri(2019)]%
        {gray2019ghost}
\bibfield{author}{\bibinfo{person}{Mary~L Gray} {and}
  \bibinfo{person}{Siddharth Suri}.} \bibinfo{year}{2019}\natexlab{}.
\newblock \bibinfo{booktitle}{\emph{Ghost work: How to stop Silicon Valley from
  building a new global underclass}}.
\newblock \bibinfo{publisher}{Eamon Dolan Books}.
\newblock


\bibitem[Hanrahan et~al\mbox{.}(2021)]%
        {hanrahan2021expertise}
\bibfield{author}{\bibinfo{person}{Benjamin~V Hanrahan}, \bibinfo{person}{Anita
  Chen}, \bibinfo{person}{JiaHua Ma}, \bibinfo{person}{Ning~F Ma},
  \bibinfo{person}{Anna Squicciarini}, {and} \bibinfo{person}{Saiph Savage}.}
  \bibinfo{year}{2021}\natexlab{}.
\newblock \showarticletitle{The Expertise Involved in Deciding which HITs are
  Worth Doing on Amazon Mechanical Turk}.
\newblock \bibinfo{journal}{\emph{Proceedings of the ACM on Human-Computer
  Interaction}} \bibinfo{volume}{5}, \bibinfo{number}{CSCW1}
  (\bibinfo{year}{2021}), \bibinfo{pages}{1--23}.
\newblock


\bibitem[Hara et~al\mbox{.}(2018)]%
        {hara2018data}
\bibfield{author}{\bibinfo{person}{Kotaro Hara}, \bibinfo{person}{Abigail
  Adams}, \bibinfo{person}{Kristy Milland}, \bibinfo{person}{Saiph Savage},
  \bibinfo{person}{Chris Callison-Burch}, {and} \bibinfo{person}{Jeffrey~P
  Bigham}.} \bibinfo{year}{2018}\natexlab{}.
\newblock \showarticletitle{A data-driven analysis of workers' earnings on
  Amazon Mechanical Turk}. In \bibinfo{booktitle}{\emph{Proceedings of the 2018
  CHI conference on human factors in computing systems}}.
  \bibinfo{pages}{1--14}.
\newblock


\bibitem[Hoffmann(2021)]%
        {termsofinclusion}
\bibfield{author}{\bibinfo{person}{Anna~Lauren Hoffmann}.}
  \bibinfo{year}{2021}\natexlab{}.
\newblock \showarticletitle{Terms of inclusion: Data, discourse, violence}.
\newblock \bibinfo{journal}{\emph{New Media \& Society}} \bibinfo{volume}{23},
  \bibinfo{number}{12} (\bibinfo{year}{2021}), \bibinfo{pages}{3539--3556}.
\newblock
\urldef\tempurl%
\url{https://doi.org/10.1177/1461444820958725}
\showDOI{\tempurl}
\showeprint{https://doi.org/10.1177/1461444820958725}


\bibitem[Holland et~al\mbox{.}(2018)]%
        {Holland2018}
\bibfield{author}{\bibinfo{person}{Sarah Holland}, \bibinfo{person}{Ahmed
  Hosny}, \bibinfo{person}{Sarah Newman}, \bibinfo{person}{Joshua Joseph},
  {and} \bibinfo{person}{Kasia Chmielinski}.} \bibinfo{year}{2018}\natexlab{}.
\newblock \showarticletitle{The dataset nutrition label: A framework to drive
  higher data quality standards}.
\newblock \bibinfo{journal}{\emph{arXiv preprint arXiv:1805.03677}}
  (\bibinfo{year}{2018}).
\newblock


\bibitem[Hutchinson et~al\mbox{.}(2021a)]%
        {hutchinson2021accountability}
\bibfield{author}{\bibinfo{person}{Ben Hutchinson}, \bibinfo{person}{Andrew
  Smart}, \bibinfo{person}{Alex Hanna}, \bibinfo{person}{Emily Denton},
  \bibinfo{person}{Christina Greer}, \bibinfo{person}{Oddur Kjartansson},
  \bibinfo{person}{Parker Barnes}, {and} \bibinfo{person}{Margaret Mitchell}.}
  \bibinfo{year}{2021}\natexlab{a}.
\newblock \showarticletitle{Towards Accountability for Machine Learning
  Datasets: Practices from Software Engineering and Infrastructure}. In
  \bibinfo{booktitle}{\emph{Proceedings of the Conference on Fairness,
  Accountability, and Transparency}}.
\newblock


\bibitem[Hutchinson et~al\mbox{.}(2021b)]%
        {Hutchinson2021}
\bibfield{author}{\bibinfo{person}{Ben Hutchinson}, \bibinfo{person}{Andrew
  Smart}, \bibinfo{person}{Alex Hanna}, \bibinfo{person}{Emily Denton},
  \bibinfo{person}{Christina Greer}, \bibinfo{person}{Oddur Kjartansson},
  \bibinfo{person}{Parker Barnes}, {and} \bibinfo{person}{Margaret Mitchell}.}
  \bibinfo{year}{2021}\natexlab{b}.
\newblock \showarticletitle{Towards Accountability for Machine Learning
  Datasets: Practices from Software Engineering and Infrastructure}. In
  \bibinfo{booktitle}{\emph{Proceedings of the 2021 ACM Conference on Fairness,
  Accountability, and Transparency}} (Virtual Event, Canada)
  \emph{(\bibinfo{series}{FAccT '21})}. \bibinfo{publisher}{Association for
  Computing Machinery}, \bibinfo{pages}{560–575}.
\newblock
\showISBNx{9781450383097}


\bibitem[Ipeirotis(2010)]%
        {ipeirotis2010demographics}
\bibfield{author}{\bibinfo{person}{Panagiotis~G Ipeirotis}.}
  \bibinfo{year}{2010}\natexlab{}.
\newblock \showarticletitle{Demographics of mechanical turk}.
\newblock  (\bibinfo{year}{2010}).
\newblock


\bibitem[Irani and Silberman(2013)]%
        {irani2013turkopticon}
\bibfield{author}{\bibinfo{person}{Lilly~C Irani} {and} \bibinfo{person}{M~Six
  Silberman}.} \bibinfo{year}{2013}\natexlab{}.
\newblock \showarticletitle{Turkopticon: Interrupting worker invisibility in
  amazon mechanical turk}. In \bibinfo{booktitle}{\emph{Proceedings of the
  SIGCHI conference on human factors in computing systems}}.
  \bibinfo{pages}{611--620}.
\newblock


\bibitem[Kaplan et~al\mbox{.}(2018)]%
        {kaplan2018striving}
\bibfield{author}{\bibinfo{person}{Toni Kaplan}, \bibinfo{person}{Susumu
  Saito}, \bibinfo{person}{Kotaro Hara}, {and} \bibinfo{person}{Jeffrey~P
  Bigham}.} \bibinfo{year}{2018}\natexlab{}.
\newblock \showarticletitle{Striving to earn more: a survey of work strategies
  and tool use among crowd workers}. In \bibinfo{booktitle}{\emph{Sixth AAAI
  Conference on Human Computation and Crowdsourcing}}.
\newblock


\bibitem[Kocsis and De~Vreede(2016)]%
        {kocsis2016towards}
\bibfield{author}{\bibinfo{person}{David Kocsis} {and}
  \bibinfo{person}{Gert~Jan De~Vreede}.} \bibinfo{year}{2016}\natexlab{}.
\newblock \showarticletitle{Towards a taxonomy of ethical considerations in
  crowdsourcing} \emph{(\bibinfo{series}{22nd Americas Conference on
  Information Systems: Surfing the IT Innovation Wave, AMCIS 2016})}.
\newblock


\bibitem[Martin et~al\mbox{.}(2014)]%
        {Martin2014}
\bibfield{author}{\bibinfo{person}{David Martin}, \bibinfo{person}{Benjamin~V.
  Hanrahan}, \bibinfo{person}{Jacki O'Neill}, {and} \bibinfo{person}{Neha
  Gupta}.} \bibinfo{year}{2014}\natexlab{}.
\newblock \showarticletitle{Being a Turker}. In
  \bibinfo{booktitle}{\emph{Proceedings of the 17th ACM Conference on Computer
  Supported Cooperative Work \& Social Computing}} (Baltimore, Maryland, USA)
  \emph{(\bibinfo{series}{CSCW '14})}. \bibinfo{publisher}{Association for
  Computing Machinery}, \bibinfo{address}{New York, NY, USA},
  \bibinfo{pages}{224–235}.
\newblock
\showISBNx{9781450325400}
\urldef\tempurl%
\url{https://doi.org/10.1145/2531602.2531663}
\showDOI{\tempurl}


\bibitem[Miceli et~al\mbox{.}(2020)]%
        {miceli2020between}
\bibfield{author}{\bibinfo{person}{Milagros Miceli}, \bibinfo{person}{Martin
  Schuessler}, {and} \bibinfo{person}{Tianling Yang}.}
  \bibinfo{year}{2020}\natexlab{}.
\newblock \showarticletitle{Between Subjectivity and Imposition: Power Dynamics
  in Data Annotation for Computer Vision}.
\newblock \bibinfo{journal}{\emph{Proc. ACM Hum.-Comput. Interact.}}
  \bibinfo{volume}{4}, \bibinfo{number}{CSCW2}, Article
  \bibinfo{articleno}{115} (\bibinfo{date}{Oct.} \bibinfo{year}{2020}),
  \bibinfo{numpages}{25}~pages.
\newblock
\urldef\tempurl%
\url{https://doi.org/10.1145/3415186}
\showDOI{\tempurl}


\bibitem[Monarch(2021)]%
        {monarch2021human}
\bibfield{author}{\bibinfo{person}{Robert~Munro Monarch}.}
  \bibinfo{year}{2021}\natexlab{}.
\newblock \bibinfo{booktitle}{\emph{Human-in-the-Loop Machine Learning: Active
  learning and annotation for human-centered AI}}.
\newblock \bibinfo{publisher}{Simon and Schuster}.
\newblock


\bibitem[Ovesdotter~Alm(2011)]%
        {alm2011subjective}
\bibfield{author}{\bibinfo{person}{Cecilia Ovesdotter~Alm}.}
  \bibinfo{year}{2011}\natexlab{}.
\newblock \showarticletitle{Subjective Natural Language Problems: Motivations,
  Applications, Characterizations, and Implications}. In
  \bibinfo{booktitle}{\emph{Proceedings of the 49th Annual Meeting of the
  Association for Computational Linguistics: Human Language Technologies}}.
  \bibinfo{publisher}{Association for Computational Linguistics},
  \bibinfo{address}{Portland, Oregon, USA}, \bibinfo{pages}{107--112}.
\newblock
\urldef\tempurl%
\url{https://aclanthology.org/P11-2019}
\showURL{%
\tempurl}


\bibitem[Patton et~al\mbox{.}(2019)]%
        {patton2019annotating}
\bibfield{author}{\bibinfo{person}{Desmond~Upton Patton},
  \bibinfo{person}{Philipp Blandfort}, \bibinfo{person}{William~R Frey},
  \bibinfo{person}{Michael~B Gaskell}, {and} \bibinfo{person}{Svebor Karaman}.}
  \bibinfo{year}{2019}\natexlab{}.
\newblock \bibinfo{title}{Annotating twitter data from vulnerable populations:
  Evaluating disagreement between domain experts and graduate student
  annotators}.
\newblock
\newblock


\bibitem[Posch et~al\mbox{.}(2018)]%
        {posch2018characterizing}
\bibfield{author}{\bibinfo{person}{Lisa Posch}, \bibinfo{person}{Arnim Bleier},
  \bibinfo{person}{Fabian Fl{\"o}ck}, {and} \bibinfo{person}{Markus
  Strohmaier}.} \bibinfo{year}{2018}\natexlab{}.
\newblock \showarticletitle{Characterizing the global crowd workforce: A
  cross-country comparison of crowdworker demographics}.
\newblock \bibinfo{journal}{\emph{arXiv preprint arXiv:1812.05948}}
  (\bibinfo{year}{2018}).
\newblock


\bibitem[Prabhakaran et~al\mbox{.}(2021)]%
        {prabhakaran2021releasing}
\bibfield{author}{\bibinfo{person}{Vinodkumar Prabhakaran},
  \bibinfo{person}{Aida~Mostafazadeh Davani}, {and} \bibinfo{person}{Mark
  Diaz}.} \bibinfo{year}{2021}\natexlab{}.
\newblock \showarticletitle{On Releasing Annotator-Level Labels and Information
  in Datasets}. In \bibinfo{booktitle}{\emph{Proceedings of the 15th Linguistic
  Annotation Workshop}}. \bibinfo{publisher}{Association for Computational
  Linguistics}, \bibinfo{address}{Virtual}.
\newblock


\bibitem[Pushkarna et~al\mbox{.}(2022)]%
        {datacards}
\bibfield{author}{\bibinfo{person}{Mahima Pushkarna}, \bibinfo{person}{Andrew
  Zaldivar}, {and} \bibinfo{person}{Oddur Kjartansson}.}
  \bibinfo{year}{2022}\natexlab{}.
\newblock \showarticletitle{Data Cards: Purposeful and Transparent Dataset
  Documentation for Responsible AI}. In \bibinfo{booktitle}{\emph{Proceedings
  of the 2021 ACM Conference on Fairness, Accountability, and Transparency}}.
\newblock


\bibitem[Quinn and Bederson(2011)]%
        {quinn2011human}
\bibfield{author}{\bibinfo{person}{Alexander~J. Quinn} {and}
  \bibinfo{person}{Benjamin~B. Bederson}.} \bibinfo{year}{2011}\natexlab{}.
\newblock \bibinfo{booktitle}{\emph{Human Computation: A Survey and Taxonomy of
  a Growing Field}}.
\newblock \bibinfo{publisher}{Association for Computing Machinery},
  \bibinfo{address}{New York, NY, USA}, \bibinfo{pages}{1403–1412}.
\newblock
\showISBNx{9781450302289}
\urldef\tempurl%
\url{https://doi.org/10.1145/1978942.1979148}
\showURL{%
\tempurl}


\bibitem[Ram\'{\i}rez et~al\mbox{.}(2021)]%
        {ramirez2021state}
\bibfield{author}{\bibinfo{person}{Jorge Ram\'{\i}rez}, \bibinfo{person}{Burcu
  Sayin}, \bibinfo{person}{Marcos Baez}, \bibinfo{person}{Fabio Casati},
  \bibinfo{person}{Luca Cernuzzi}, \bibinfo{person}{Boualem Benatallah}, {and}
  \bibinfo{person}{Gianluca Demartini}.} \bibinfo{year}{2021}\natexlab{}.
\newblock \showarticletitle{On the State of Reporting in Crowdsourcing
  Experiments and a Checklist to Aid Current Practices}.
\newblock \bibinfo{journal}{\emph{Proc. ACM Hum.-Comput. Interact.}}
  \bibinfo{volume}{5}, \bibinfo{number}{CSCW2}, Article
  \bibinfo{articleno}{387} (\bibinfo{date}{oct} \bibinfo{year}{2021}),
  \bibinfo{numpages}{34}~pages.
\newblock
\urldef\tempurl%
\url{https://doi.org/10.1145/3479531}
\showDOI{\tempurl}


\bibitem[Roberts(2016)]%
        {roberts2016digital}
\bibfield{author}{\bibinfo{person}{Sarah~T Roberts}.}
  \bibinfo{year}{2016}\natexlab{}.
\newblock \showarticletitle{Digital refuse: Canadian garbage, commercial
  content moderation and the global circulation of social media’s waste}.
\newblock \bibinfo{journal}{\emph{Wi: Journal of Mobile Media}}
  \bibinfo{volume}{10}, \bibinfo{number}{1} (\bibinfo{year}{2016}),
  \bibinfo{pages}{1--18}.
\newblock


\bibitem[Rostamzadeh et~al\mbox{.}(2021)]%
        {healthsheets}
\bibfield{author}{\bibinfo{person}{Negar Rostamzadeh},
  \bibinfo{person}{Subhrajit Roy}, \bibinfo{person}{Diana Mincu},
  \bibinfo{person}{Andrew Smart}, \bibinfo{person}{Lauren Wilcox},
  \bibinfo{person}{Mahima Pushkarna}, \bibinfo{person}{Razvan Amironesei},
  \bibinfo{person}{Jessica Schrouff}, \bibinfo{person}{Madeleine Elish},
  \bibinfo{person}{Nyalleng Moorosi}, \bibinfo{person}{Berk Ustun},
  \bibinfo{person}{Noah Broesti}, {and} \bibinfo{person}{Katherine Heller}.}
  \bibinfo{year}{2021}\natexlab{}.
\newblock \showarticletitle{Specialized Healthsheet for Healthcare Datasets}.
  In \bibinfo{booktitle}{\emph{Machine Learning for Health (ML4H)}}.
\newblock


\bibitem[Sabou et~al\mbox{.}(2014)]%
        {sabou2014corpus}
\bibfield{author}{\bibinfo{person}{Reka~Marta Sabou}, \bibinfo{person}{Kalina
  Bontcheva}, \bibinfo{person}{Leon Derczynski}, {and} \bibinfo{person}{A.
  Scharl}.} \bibinfo{year}{2014}\natexlab{}.
\newblock \showarticletitle{Corpus Annotation through Crowdsourcing:Towards
  Best Practice Guidelines}. In \bibinfo{booktitle}{\emph{Proceedings of the
  Ninth International Conference on Language Resources and Evaluation
  (LREC'14)}}. \bibinfo{publisher}{European Language Resources Association
  (ELRA)}, \bibinfo{address}{Reykjavik, Iceland}.
\newblock
\showISBNx{978-2-9517408-8-4}


\bibitem[Salehi et~al\mbox{.}(2015)]%
        {Salehi2015}
\bibfield{author}{\bibinfo{person}{Niloufar Salehi}, \bibinfo{person}{Lilly~C.
  Irani}, \bibinfo{person}{Michael~S. Bernstein}, \bibinfo{person}{Ali
  Alkhatib}, \bibinfo{person}{Eva Ogbe}, \bibinfo{person}{Kristy Milland},
  {and} \bibinfo{person}{Clickhappier}.} \bibinfo{year}{2015}\natexlab{}.
\newblock \bibinfo{booktitle}{\emph{We Are Dynamo: Overcoming Stalling and
  Friction in Collective Action for Crowd Workers}}.
\newblock \bibinfo{publisher}{Association for Computing Machinery},
  \bibinfo{address}{New York, NY, USA}, \bibinfo{pages}{1621–1630}.
\newblock
\showISBNx{9781450331456}
\urldef\tempurl%
\url{https://doi.org/10.1145/2702123.2702508}
\showURL{%
\tempurl}


\bibitem[Sambasivan et~al\mbox{.}(2021)]%
        {sambasivan2021re}
\bibfield{author}{\bibinfo{person}{Nithya Sambasivan}, \bibinfo{person}{Erin
  Arnesen}, \bibinfo{person}{Ben Hutchinson}, \bibinfo{person}{Tulsee Doshi},
  {and} \bibinfo{person}{Vinodkumar Prabhakaran}.}
  \bibinfo{year}{2021}\natexlab{}.
\newblock \showarticletitle{Re-Imagining Algorithmic Fairness in India and
  Beyond}. In \bibinfo{booktitle}{\emph{Proceedings of the 2021 ACM Conference
  on Fairness, Accountability, and Transparency}} (Virtual Event, Canada)
  \emph{(\bibinfo{series}{FAccT '21})}. \bibinfo{publisher}{Association for
  Computing Machinery}, \bibinfo{address}{New York, NY, USA},
  \bibinfo{pages}{315–328}.
\newblock
\showISBNx{9781450383097}
\urldef\tempurl%
\url{https://doi.org/10.1145/3442188.3445896}
\showDOI{\tempurl}


\bibitem[Scheuerman et~al\mbox{.}(2021)]%
        {scheuerman2021}
\bibfield{author}{\bibinfo{person}{Morgan~Klaus Scheuerman},
  \bibinfo{person}{Emily Denton}, {and} \bibinfo{person}{Alex Hanna}.}
  \bibinfo{year}{2021}\natexlab{}.
\newblock \showarticletitle{Do Datasets Have Politics? Disciplinary Values in
  Computer Vision Dataset Development}.
\newblock \bibinfo{journal}{\emph{Computer Supported Cooperative Work (CSCW)}}
  (\bibinfo{year}{2021}).
\newblock


\bibitem[Schlagwein et~al\mbox{.}(2019)]%
        {schlagwein2019ethical}
\bibfield{author}{\bibinfo{person}{Daniel Schlagwein},
  \bibinfo{person}{Dubravka Cecez-Kecmanovic}, {and} \bibinfo{person}{Benjamin
  Hanckel}.} \bibinfo{year}{2019}\natexlab{}.
\newblock \showarticletitle{Ethical norms and issues in crowdsourcing
  practices: A Habermasian analysis}.
\newblock \bibinfo{journal}{\emph{Information Systems Journal}}
  \bibinfo{volume}{29}, \bibinfo{number}{4} (\bibinfo{year}{2019}),
  \bibinfo{pages}{811--837}.
\newblock
\urldef\tempurl%
\url{https://doi.org/10.1111/isj.12227}
\showDOI{\tempurl}
\showeprint{https://onlinelibrary.wiley.com/doi/pdf/10.1111/isj.12227}


\bibitem[Seamster and Charron-Chénier(2017)]%
        {seamster2017}
\bibfield{author}{\bibinfo{person}{Louise Seamster} {and}
  \bibinfo{person}{Raphaël Charron-Chénier}.}
  \bibinfo{year}{2017}\natexlab{}.
\newblock \showarticletitle{Predatory Inclusion and Education Debt: Rethinking
  the Racial Wealth Gap}.
\newblock \bibinfo{journal}{\emph{Social Currents}} \bibinfo{volume}{4},
  \bibinfo{number}{3} (\bibinfo{year}{2017}), \bibinfo{pages}{199--207}.
\newblock
\urldef\tempurl%
\url{https://doi.org/10.1177/2329496516686620}
\showDOI{\tempurl}
\showeprint{https://doi.org/10.1177/2329496516686620}


\bibitem[Selbst et~al\mbox{.}(2019)]%
        {selbst2018fairness}
\bibfield{author}{\bibinfo{person}{Andrew~D. Selbst}, \bibinfo{person}{Danah
  Boyd}, \bibinfo{person}{Sorelle~A. Friedler}, \bibinfo{person}{Suresh
  Venkatasubramanian}, {and} \bibinfo{person}{Janet Vertesi}.}
  \bibinfo{year}{2019}\natexlab{}.
\newblock \showarticletitle{Fairness and Abstraction in Sociotechnical
  Systems}. In \bibinfo{booktitle}{\emph{Proceedings of the Conference on
  Fairness, Accountability, and Transparency}} (Atlanta, GA, USA)
  \emph{(\bibinfo{series}{FAT* '19})}. \bibinfo{publisher}{Association for
  Computing Machinery}, \bibinfo{address}{New York, NY, USA},
  \bibinfo{pages}{59–68}.
\newblock
\showISBNx{9781450361255}
\urldef\tempurl%
\url{https://doi.org/10.1145/3287560.3287598}
\showDOI{\tempurl}


\bibitem[Semuels(2018)]%
        {semuels2018internet}
\bibfield{author}{\bibinfo{person}{Alana Semuels}.}
  \bibinfo{year}{2018}\natexlab{}.
\newblock \showarticletitle{The internet is enabling a new kind of poorly paid
  hell}.
\newblock \bibinfo{journal}{\emph{The Atlantic}}  \bibinfo{volume}{23}
  (\bibinfo{year}{2018}).
\newblock


\bibitem[Sen et~al\mbox{.}(2015)]%
        {sen2015turkers}
\bibfield{author}{\bibinfo{person}{Shilad Sen}, \bibinfo{person}{Margaret~E.
  Giesel}, \bibinfo{person}{Rebecca Gold}, \bibinfo{person}{Benjamin Hillmann},
  \bibinfo{person}{Matt Lesicko}, \bibinfo{person}{Samuel Naden},
  \bibinfo{person}{Jesse Russell}, \bibinfo{person}{Zixiao~(Ken) Wang}, {and}
  \bibinfo{person}{Brent Hecht}.} \bibinfo{year}{2015}\natexlab{}.
\newblock \showarticletitle{Turkers, Scholars, "Arafat" and "Peace": Cultural
  Communities and Algorithmic Gold Standards}. In
  \bibinfo{booktitle}{\emph{Proceedings of the 18th ACM Conference on Computer
  Supported Cooperative Work \& Social Computing}} (Vancouver, BC, Canada)
  \emph{(\bibinfo{series}{CSCW '15})}. \bibinfo{publisher}{Association for
  Computing Machinery}, \bibinfo{address}{New York, NY, USA},
  \bibinfo{pages}{826–838}.
\newblock
\showISBNx{9781450329224}
\urldef\tempurl%
\url{https://doi.org/10.1145/2675133.2675285}
\showDOI{\tempurl}


\bibitem[Shmueli et~al\mbox{.}(2021)]%
        {shmueli2021beyond}
\bibfield{author}{\bibinfo{person}{Boaz Shmueli}, \bibinfo{person}{Jan Fell},
  \bibinfo{person}{Soumya Ray}, {and} \bibinfo{person}{Lun-Wei Ku}.}
  \bibinfo{year}{2021}\natexlab{}.
\newblock \showarticletitle{Beyond Fair Pay: Ethical Implications of NLP
  Crowdsourcing}.
\newblock \bibinfo{journal}{\emph{arXiv preprint arXiv:2104.10097}}
  (\bibinfo{year}{2021}).
\newblock


\bibitem[Srinivasan et~al\mbox{.}(2021)]%
        {artsheets}
\bibfield{author}{\bibinfo{person}{Ramya~Malur Srinivasan},
  \bibinfo{person}{Emily Denton}, \bibinfo{person}{Jordan~Jennifer Famularo},
  \bibinfo{person}{Negar Rostamzadeh}, \bibinfo{person}{Fernando Diaz}, {and}
  \bibinfo{person}{Beth Coleman}.} \bibinfo{year}{2021}\natexlab{}.
\newblock \showarticletitle{Artsheets for Art Datasets}. In
  \bibinfo{booktitle}{\emph{Proceedings of Neural Information Processing
  Systems (NeurIPS), Datasets \& Benchmarks Track}}.
\newblock
\urldef\tempurl%
\url{https://openreview.net/pdf?id=K7ke_GZ_6N}
\showURL{%
\tempurl}


\bibitem[Toxtli et~al\mbox{.}(2021)]%
        {Toxtli2021}
\bibfield{author}{\bibinfo{person}{Carlos Toxtli}, \bibinfo{person}{Siddharth
  Suri}, {and} \bibinfo{person}{Saiph Savage}.}
  \bibinfo{year}{2021}\natexlab{}.
\newblock \showarticletitle{Quantifying the Invisible Labor in Crowd Work}.
\newblock \bibinfo{journal}{\emph{Proc. ACM Hum.-Comput. Interact.}}
  \bibinfo{volume}{5}, \bibinfo{number}{CSCW2} (\bibinfo{year}{2021}).
\newblock


\bibitem[Vakharia and Lease(2015)]%
        {vakharia2015beyond}
\bibfield{author}{\bibinfo{person}{Donna Vakharia} {and}
  \bibinfo{person}{Matthew Lease}.} \bibinfo{year}{2015}\natexlab{}.
\newblock \showarticletitle{Beyond Mechanical Turk: An analysis of paid crowd
  work platforms}.
\newblock \bibinfo{journal}{\emph{Proceedings of the iConference}}
  (\bibinfo{year}{2015}), \bibinfo{pages}{1--17}.
\newblock


\bibitem[Vogels(2021)]%
        {vogels2021state}
\bibfield{author}{\bibinfo{person}{Emily Vogels}.}
  \bibinfo{year}{2021}\natexlab{}.
\newblock \showarticletitle{The state of online harassment}.
\newblock \bibinfo{journal}{\emph{Pew Research Center}} (\bibinfo{year}{2021}).
\newblock


\bibitem[Waseem(2016)]%
        {waseem2016racist}
\bibfield{author}{\bibinfo{person}{Zeerak Waseem}.}
  \bibinfo{year}{2016}\natexlab{}.
\newblock \showarticletitle{Are You a Racist or Am {I} Seeing Things? Annotator
  Influence on Hate Speech Detection on {T}witter}. In
  \bibinfo{booktitle}{\emph{Proceedings of the First Workshop on {NLP} and
  Computational Social Science}}. \bibinfo{publisher}{Association for
  Computational Linguistics}, \bibinfo{address}{Austin, Texas},
  \bibinfo{pages}{138--142}.
\newblock
\urldef\tempurl%
\url{https://doi.org/10.18653/v1/W16-5618}
\showDOI{\tempurl}


\end{thebibliography}

\end{document}